\documentstyle[12pt]{article}
\begin{document}
\newcommand{\eqn}[1]{eq.(\ref{#1})}

\renewcommand{\section}[1]{\addtocounter{section}{1}
\vspace{5mm} \par \noindent
  {\bf \thesection . #1}\setcounter{subsection}{0}
  \par
   \vspace{2mm} } 
\newcommand{\sectionsub}[1]{\addtocounter{section}{1}
\vspace{5mm} \par \noindent
  {\bf \thesection . #1}\setcounter{subsection}{0}\par}
\renewcommand{\subsection}[1]{\addtocounter{subsection}{1}
\vspace{2.5mm}\par\noindent {\em \thesubsection . #1}\par
 \vspace{0.5mm} }
\renewcommand{\thebibliography}[1]{ {\vspace{5mm}\par \noindent{\bf
References}\par \vspace{2mm}}
\list
 {\arabic{enumi}.}{\settowidth\labelwidth{[#1]}\leftmargin\labelwidth
 \advance\leftmargin\labelsep\addtolength{\topsep}{-4em}
 \usecounter{enumi}}
 \def\newblock{\hskip .11em plus .33em minus .07em}
 \sloppy\clubpenalty4000\widowpenalty4000
 \sfcode`\.=1000\relax \setlength{\itemsep}{-0.4em} }

\newcommand{\eq}[1]{eq.(\ref{#1})}
\newcommand{\Eq}[1]{Eq.(\ref{#1})}
\newcommand{\ben}{\begin{equation}}
\newcommand{\een}{\end{equation}}
\newcommand{\bea}{\begin{eqnarray}}
\newcommand{\eea}{\end{eqnarray}}
\newcommand{\bear}{\begin{array}}
\newcommand{\enar}{\end{array}}
\newcommand{\bdm}{\begin{displaymath}}
\newcommand{\edm}{\end{displaymath}}
\newcommand{\nn}{\nonumber \\ }
\newcommand{\binomial}[2]{\left (\begin{array}{c} {#1}\\ {#2} \end{array}
\right )}
\newcommand{\hf}{\frac{1}{2}}
\newcommand{\thf}{\frac{3}{2}}
\newcommand{\vx}{\vec{x}}
\newcommand{\vk}{\vec{k}}
\newcommand{\vn}{\vec{\nabla}}
\newcommand{\ve}{\vec{e}}
\newcommand{\vep}{\vec{e}\mbox{\hspace{1mm}}'}
\newcommand{\La}{{\cal L}}
\newcommand{\Ha}{{\cal H}}
\newcommand{\pa}{\partial}
\newcommand{\del}{\delta}
\newcommand{\eps}{\epsilon}
\newcommand{\al}{\alpha}
\newcommand{\fslu}{\not\mbox{\hspace{-1.5mm}}}
\newcommand{\fsll}{\not\mbox{\hspace{-.5mm}}}
\newcommand{\D}{{\cal D}}
\newcommand{\N}{{\cal N}}
\newcommand{\tb}{\overline{t}}
\newcommand{\qh}{\hat{q}}
\newcommand{\phih}{\hat{\phi}}
\newcommand{\g}{\gamma}
\newcommand{\G}{\Gamma}
\newcommand{\R}{\mbox{\hspace{.04mm}\rule{0.2mm}{2.8mm}\hspace{-1.5mm} R}}
\newcommand{\C}{\mbox{\hspace{1.24mm}\rule{0.2mm}{2.5mm}\hspace{-2.7mm} C}}
\newcommand{\Q}{\mbox{\hspace{1.24mm}\rule{0.2mm}{2.7mm}\hspace{-2.7mm} Q}}
\newcommand{\Z}{\mbox{$Z\hspace{-2mm}Z$}}
\newcommand{\Nat}{\mbox{\hspace{.04mm}\rule{0.2mm}{2.8mm}\hspace{-1.5mm} N}}
\newcommand{\gb}{\overline{\gamma}}
\newcommand{\ztb}{\overline{\zeta}}
\newcommand{\wb}{\overline{w}}
\newcommand{\zt}{\zeta}
\newcommand{\pss}{\psi^*}
\newcommand{\zh}{{\sf Z}+\hf}
\newcommand{\zs}{{\sf Z}}
\newcommand{\tq}{\tilde{q}}
\newcommand{\br}{\langle}
\newcommand{\kt}{\rangle}
\newcommand{\bra}[1]{\langle {#1}|}
\newcommand{\ket}[1]{|{#1}\rangle}
\newcommand{\vac}{\ket{0}}
\newcommand{\kad}{\ket{0,D}}
\newcommand{\bad}{\bra{0,D}}
\newcommand{\dzp}{\frac{dz}{2\pi i}}
\newcommand{\dtp}[1]{\frac{d{#1}}{2\pi i}}
\newcommand{\Asf}{{\sf A}}
\newcommand{\Bsf}{{\sf B}}
\newcommand{\Csf}{{\sf C}}
\newcommand{\ps}{\psi}
\newcommand{\psb}{\overline{\psi}}
\newcommand{\etb}{\overline{\eta}}
\newcommand{\Phc}{\Phi^{Cl}}
\newcommand{\dz}{\frac{dz}{2\pi i}}
\newcommand{\vs}{\vspace}
\renewcommand{\Im}{{\rm Im}\,}
\newcommand{\NP}[1]{Nucl.\ Phys.\ {\bf #1}}
\newcommand{\PL}[1]{Phys.\ Lett.\ {\bf #1}}
\newcommand{\CMP}[1]{Commun.\ Math.\ Phys.\ {\bf #1}}
\newcommand{\PR}[1]{Phys.\ Rev.\ {\bf #1}}
\newcommand{\PRL}[1]{Phys.\ Rev.\ Lett.\ {\bf #1}}
\newcommand{\PTP}[1]{Prog.\ Theor.\ Phys.\ {\bf #1}}
\newcommand{\PTPS}[1]{Prog.\ Theor.\ Phys.\ Suppl.\ {\bf #1}}
\newcommand{\MPL}[1]{Mod.\ Phys.\ Lett.\ {\bf #1}}
\newcommand{\IJMP}[1]{Int.\ Jour.\ Mod.\ Phys.\ {\bf #1}}
\newcommand{\IM}[1]{Invent.\ Math.\ {\bf #1}}
\newcommand{\SJNP}[1]{Sov. J. Nucl. Phys.\ {\bf #1}}

\begin{flushright}
NBI-HE-95-33 \\
hep-th/9510059
\end{flushright}

\vspace{4mm}
\begin{center}
{\bf CONFORMAL BLOCKS FOR ADMISSIBLE}\\
{\bf REPRESENTATIONS IN SL(2) CURRENT ALGEBRA}
\footnote{Talk presented by J. Rasmussen}
\vspace{1.4cm}

J.L.~Petersen, J.~Rasmussen and M.~Yu \\
{\em The Niels Bohr Institute} \\
{\em Blegdamsvej 17, DK-2100 Copenhagen {\O}, Denmark} \\
\end{center}
\centerline{ABSTRACT}
\vspace{- 4 mm}  
\begin{quote}\small
A review is presented of the recently obtained expressions for conformal blocks
for {\it admissible} representations in $SL(2)$ current algebra based on
the Wakimoto free field construction. In this
realization one needs to introduce a second screening charge, one which
depends on fractional powers of free fields. The techniques necessary to deal
with these complications are developed, and explicit general integral
representations for conformal blocks on the sphere are provided. The fusion
rules are discussed and as a check it is verified that the conformal blocks
satisfy the Knizhnik-Zamolodchikov equations.
\end{quote}
\addtocounter{section}{1}
\par \noindent
  {\bf \thesection . Introduction}
  \par
   \vspace{2mm} 

\noindent
In refs. \cite{HY}-\cite{AGSY}
H.-L. Hu and M. Yu and O. Aharony et al have proposed an
equivalence between the coupling of usual conformal minimal matter to 2-$d$
gravity using Hamiltonian reduction and the twisted $SL(2)/SL(2)$ WZNW models.
Their discussions cover comparisons of field contents and cohomologies.
These works constitute our motivation for studying $N$-point correlation
functions of 2-$d$ conformal WZNW theories based on affine
$\widehat{SL}(2)_k$
since the cases of admissible representations \cite{KW}
are the ones relevant for treating the minimal matter. Much
attention has been paid to the $N$ point correlators either by applying
the Wakimoto free field realization \cite{Wak}, from which results have been
given in refs. \cite{BF,ATY,FGPP,D90},
or by solving the Knizhnik-Zamolodchikov (KZ) equations
\cite{KZ}, from which results have been giving in e.g. refs.
\cite{KZ,FZ,A,CF,SV,FGPP,FIM}. The results
are quite complete as far as unitary, integrable representations are concerned,
but have appeared incomplete for the general case including the admissible
representations. In ref. \cite{PRY1} we have found complete integral
expressions in the case of admissible representations , based on the free
field construction, for exactly the conformal blocks relevant to the above
applications.

In general the WZNW theory is characterized by the level, $k$, or equivalently
by $t=k+2$ (for $\widehat{SL}(2)_k$). Then  degenerate primary fields
exist for representations characterized by spins, $j_{r,s}$, given by
\cite{KK,MFF}
\ben
2j_{r,s}+1=r-st
\een
with $r,s$ integers. However, previous applications of the free field Wakimoto
realization can be characterized as
being fully complete only for the case, $s=0$, which is the full case only for
integrable representations.
The reason for this restriction is fairly natural, since (see sect. 2)
the screening charge usually employed in the free field realization is capable
of screening just such primary fields. In fact, a possible second screening
operator, capable of screening the general case, was proposed by Bershadsky
and Ooguri \cite{BO}, but since it involved fractional powers of the free
ghost fields, discussions on its interpretation have been only partly
successful
\cite{D90,FGPP}.
In ref. \cite{PRY1} we have overcome this difficulty by showing
how the techniques
of fractional calculus \cite{MR} naturally provide a solution.
It should be mentioned that the authors of ref. \cite{FIM} have found a
different
class of solutions to the KZ equations than the class belonging to the
integrable representations, but as we discuss in \cite{PRY1} their solutions
only have a rather small overlap with ours and are insufficient for solving
the theory.

We refer
to \cite{PRY1} for more details.

\section{Notation. Introduction of fractional calculus}
\noindent
The $\widehat{SL}(2)_k$ affine
current algebra may be written as
\bea
J^+(z)J^-(w)&=&\frac{2}{z-w}J^3(w)+\frac{k}{(z-w)^2}\nn
J^3(z)J^\pm(w)&=&\pm\frac{1}{z-w}J^\pm(w)\nn
J^3(z)J^3(w)&=&\frac{k/2}{(z-w)^2}
\eea
when we only consider one chirality of the fields.
The Wakimoto realization \cite{Wak} is based on the free scalar field,
$\varphi(z)$, and bosonic ghost fields, $(\beta(z),\gamma(z))$,
of dimensions $(1,0)$ which we take to have the following contractions
\ben
\varphi(z)\varphi(w)=\log(z-w), \ \ \ \beta(z)\gamma(w)=\frac{1}{z-w}
\een
The currents are represented as
\bea
J^+(z)&=&\beta(z)\nn
J^3(z)&=&-:\gamma\beta:(z)-\sqrt{t/2}\pa\varphi(z)\nn
J^-(z)&=&-:\gamma^2\beta:(z)+k\pa\gamma(z)-\sqrt{2t}\gamma\pa\varphi(z)\nn
t&\equiv&k+2\neq 0
\label{wakimoto}
\eea
while the Sugawara energy momentum tensor is obtained as
\ben
T(z)=:\beta\pa\gamma:(z)+\frac{1}{2}:\pa\varphi\pa\varphi:(z)+
\frac{1}{\sqrt{2t}}\pa^2\varphi(z)
\een
with central charge
\ben
c=\frac{3k}{k+2}
\een
Very conveniently one can collect the primary fields in multiplets
$\phi_j(w,x)$ (cf. \cite{FZ}) parametrized
by a variable $x$ which keeps track of the $SL(2)$ representation, $j$
\ben
J^a(z)\phi_j(w,x)=\frac{1}{z-w}J_0^a(w)\phi_j(w,x)
\label{prim}
\een
where the $SL(2)$ representation is provided by the differential operators
\bea
J_0^a(z)\phi_j(z,x)&=&[J_0^a, \phi_j(z,x)]=D_x^a \phi_j(z,x)\nn
D_x^+&=&-x^2\pa_x+2xj\nn
D_x^3&=&-x\pa_x+j\nn
D_x^-&=&\pa_x
\eea
It is a matter of direct verification to check that
\ben
\phi_j(z,x)=(1+\gamma(z)x)^{2j}:e^{-j\sqrt{2/t}\varphi(z)}:
\label{pfd}
\een
is a primary field as defined above.
In the case of admissible representations
\bea
t&=&p/q\nn
2j_i+1&=&r_i-s_it\nn
1&\leq& r_i\leq p-1 \nn
0&\leq& s_i\leq q-1
\label{adm}
\eea
where $(p,q)=1$ and $p,q\in\Nat$, one needs {\it two} screening currents
\bea
S_1(z)&=&\beta(z)e^{+\sqrt{2/t}\varphi(z)}\nn
S_2(z)&=&\beta(z)^{-t}e^{-\sqrt{2t}\varphi(z)}
\label{screen}
\eea
The labelling of the $j$'s in (\ref{adm})
refers to the $N$ primary fields in the correlators.
Screening currents are dimension 1 fields and have total
derivatives in the OPE's not only with the energy momentum tensor
but also with the
affine currents. This ensures that the screening charges (integrated
screening currents) can be inserted into the correlators without
spoiling the affine Ward identities.

The appearance of the fractional ghost field (\ref{screen}) leads to the
necessity of generalizing the usual Wick contractions which may be
written
\bea
\beta(z)^n F(\gamma(w))&=&:(\beta(z)+\frac{1}{z-w}\pa_{\gamma(w)})^n
F(\gamma(w)
):\nn
\gamma(z)^n F(\beta(w))&=&:(\gamma(z)-\frac{1}{z-w}\pa_{\beta(w)})^n F(\beta(w)
):
\label{bgcon}
\eea
Our proposal to deal with entities like $\beta(z)^{-t}$ consists in the
following generalization of (\ref{bgcon})
\ben
G(\beta(z)) F(\gamma(w))=:G(\beta(z)+\frac{1}{z-w}\pa_{\gamma(w)})F(\gamma(w)):
\label{funcbgcon}
\een
To be able to perform these we need to know how to expand the different
expressions, for instance
\ben
(1+\gamma(z)x)_{(\alpha)}^{2j}=
\sum_{n\in \Z}\binomial{2j}{n+\alpha}(\gamma(z)x)^{n+\alpha}
\een
which appears as the ghost content of the primary field. The different
choices of asymptotic expansions have been labelled by a parameter $\al$.
When deciding on what expansions to adopt, we use the criterion that
{\it after} all Wick contractions have been performed the powers, which
are then inside normal ordering signs, are non-negative integers. Then the
resulting terms have an obvious interpretation when sandwiched between
bra and ket states.

For non-unitary representations as the admissible ones, $2j$ is not
necessarily integer and we see the need for fractional
calculus. We use
\ben
\pa^a_x x^b=\frac{\Gamma(b+1)}{\Gamma(b-a+1)}x^{b-a}
\een
as the basic definition of fractional differentiation and
\ben
D^a\exp (x)=\sum_{n\in\Z}\frac{1}{\Gamma(n-a+1)}x^{n-a}, \ \ a\in\C
\een
is then an example of an asymptotic expansion of the exponential function.

Before considering correlators let us briefly go through our notation
for mode expansions, vacuum states etc.
Using the mode expansions
\bea
j(z)&=&-:\gamma(z)\beta(z): =+\pa\phi(z)\nn
\varphi(z)\varphi(z')&=&+\log(z-z')\nn
\phi(z)\phi(z')&=&-\log(z-z')\nn
\varphi(z)&=&q_\varphi +a_0\log z +\sum_{n\neq 0}\frac{a_n}{-n}z^{-n}\nn
\phi(z)&=&q_\phi+j_0\log z+\sum_{n\neq 0}\frac{j_n}{-n}z^{-n}\nn
{[}a_0,q_\varphi{]}&=&+1\nn
{[}j_0,q_\phi{]}&=&-1
\eea
the dual vacuum state $\bra{0}$ is defined by
\ben
\bra{0}=\bra{sl_2}e^{-q_\phi}e^{\sqrt{2/t}q_\varphi}
\een
where $\bra{sl_2}$ is the usual $SL(2)$ invariant bra vacuum, while the ket
vacuum, $\ket{0}$, is identical to the $SL(2)$ invariant ket vacuum
$\ket{sl_2}$.
{}From these states we construct dual bra states of {\it lowest} $SL(2)$ weight
\ben
\bra{j}=\bra{0}e^{j\sqrt{2/t}q_\varphi}
\een
and similarly the {\it highest} weight ket state
\ben
\ket{j}=e^{-j\sqrt{2/t}q_\varphi}\ket{0}
\een
They are normalized such that
\ben
\bra{j}j\kt=1
\een

\section{Three point functions and fusion rules}
\noindent
Let us now consider the evaluation of the (chiral) three point function
\ben
\bra{j_3}\phi_{j_2}(z,x)\ket{j_1}
\een
Using the
free field realization (\ref{pfd}) of $\phi_{j_2}(z,x)$ the three point
function may be evaluated only provided the "momentas" or "charges" may be
screened away in the standard way \cite{DF}, and correspondingly
$\phi_{j_2}(z,x)$ is replaced by the intertwining field,
$(\phi_{j_2}(z,x))_{j_1}^{j_3}$,
which maps a $j_1$ highest weight module into a $j_3$ highest weight module.
Following Felder \cite{F,BF}, but using the two
screening charges in (\ref{screen}) instead, we are led to consider
the intertwining field
\bea
(\phi_{j_2}(z,x))_{j_1}^{j_3}&=&\oint\prod_{j=1}^s
\dtp{v_j}\prod_{i=1}^r\dtp{u_i}\phi_{j_2}(z,x)P(u_1,...,u_r;v_1,...,v_s)\nn
P(u_1,...,u_r;v_1,...,v_s)&=&\prod_{j=1}^s\beta^{-t}(v_j)
e^{-\sqrt{2t}\varphi(v_j)}\prod_{i=1}^r\beta(u_i)e^{\sqrt{2/t}\varphi(u_i)}
\label{intertw}
\eea
This requires that
\ben
j_1+j_2-j_3=r-st
\label{3rs}
\een
with $r$ and $s$ non-negative integers.
It is trivial using well known techniques to perform the $\varphi$ part of the
Wick contractions. Hence we concentrate on explaining how to perform the
ghost part. The calculations rely on the following two lemmas\\
{\bf Lemma 1}
\ben
(1+\gamma(z)x)^{2j}=\Gamma(2j+1)\oint_0\dtp{u}\frac{1}{u}(u^{-1}D)^{-2j}
\exp{[}(1+\gamma(z)x)/u{]}
\een
This is an
almost trivial integral representation of the ghost part of the primary
field.\\
{\bf Lemma 2}
\ben
\beta^a(w)\exp{[}(1+\gamma(z)x)/u{]}=:(\beta(w)+\frac{x/u}{w-z})^aD^a
\exp{[}(1+\gamma(z)x)/u{]}:
\label{lemma}
\een
This second lemma tells us how to perform the contraction between a fractional
$\beta$ field and the $\gamma$ content of the integral
representation in the first
lemma. Now it is straightforward to obtain the total ghost part of the
contractions. After all contractions have been carried out, the sandwiching
between the dual bra and the ket results in effectively putting
$\beta=\gamma=0$ because of the normal ordering.
In the case of admissible representations we reach an integral expression for
the three point function $W_3$
\bea
W_3&=&\frac{\Gamma(2j_2+1)}{\Gamma(2j_2-r+st+1)}\nn
&\cdot&\oint\prod_{i=1}^r\dtp{u_i}\prod_{j=1}^s\dtp{v_j}\prod_{i_1<i_2}
(u_{i_1}-u_{i_2})^{2/t}\prod_{j_1<j_2}(v_{j_1}-v_{j_2})^{2t}
\prod_{i,j}(u_i-v_j)^{-2}\nn
&\cdot&\prod_{i=1}^ru_i^{(1-r_1)/t+s_1}(1-u_i)^{(1-r_2)/t+s_2-1}
\prod_{j=1}^sv_j^{r_1-1-s_1t}(1-v_j)^{r_2-1-(s_2-1)t}
\eea
Finally the $u$ and $v$ integrations
around the Felder contours are of the Dotsenko-Fateev \cite{DF} form and may be
performed explicitly
\bea
W_3&=&\frac{\Gamma(2j_2+1)}{\Gamma(j_2+j_3-j_1+1)}
e^{i\pi r(r+1-2r_1)/t}e^{i\pi ts(s-1-2s_1)}t^{rs} \nn
&\cdot&\prod_{j=1}^r\frac{(1-e^{2\pi i(r_1-j)/t})(1-e^{2\pi i j/t})}
{1-e^{2\pi i/t}}
\prod_{j=1}^s\frac{(1-e^{2\pi it(s_1+1-j)})(1-e^{2\pi itj})}
{1-e^{2\pi it}}\nn
&\cdot&\prod_{i=1}^r\frac{\Gamma(i/t)}{\Gamma(1/t)}\prod_{i=1}^s
\frac{\Gamma(it-s)}{\Gamma (t)}\nn
&\cdot&\prod_{i=0}^{r-1}\frac{\Gamma(s_1+1+(1-r_1+i)/t)\Gamma(s_2+(1-r_2+i)/t)}
{\Gamma(s_1+s_2+1-2s+(r-r_1-r_2+i+1)/t)}\nn
&\cdot&\prod_{i=0}^{s-1}\frac{\Gamma(r_1-r+(i-s_1)t)\Gamma(r_2-r+(1-s_2+i)t)}
{\Gamma(r_1-r+r_2+(s-s_1-s_2+i)t)}
\eea
The analysis of this expression in terms of fusion rules is standard \cite{F}.
The result may be written as follows
\bea
1+|r_1-r_2|\leq&r_3&\leq p-1-|r_1+r_2-p|\nn
|s_1-s_2|\leq&s_3&\leq q-1-|s_1+s_2-q+1|
\label{FI}
\eea
The first line of these fusion rules is
well known for the case, $q=1$, of integrable
representations, and it was obtained in the general case in \cite{BF}. The
second was obtained by Awata and Yamada \cite{AY}
by considering the conditions for decoupling of null-states,
and by Feigin and Malikov \cite{FM} by cohomological methods. In
addition these authors provide a fusion rule ((II) for \cite{AY}, (I) for
\cite{FM}), which we do not get in the
free field realization. We do not know if there exist conformal field
theories with non-vanishing couplings respecting those.

\section{$N$ point functions}
\noindent
We wish to evaluate the conformal block
\ben
W_N=\bra{j_N}{[}\phi_{j_{N-1}}(z_{N-1},x_{N-1}){]}^{j_N}_{\iota_{N-2}}
...{[}\phi_{j_{n}}(z_{n},x_n){]}^{\iota_{n}}_{\iota_{n-1}}...
{[}\phi_{j_{2}}(z_2,x_2){]}^{\iota_2}_{j_1}\ket{j_1}
\een
{}From the pictorial version
\ben
\begin{picture}(300,70)
\put(0,0){$j_N$}  \put(10,10){\line(1,0){50}}  \put(55,0){$\iota_{N-2}$}
\put(35,10){\line(0,1){40}}  \put(35,55){$j_{N-1}$}
\put(110,10){\line(1,0){50}}
\put(110,0){$\iota_n$}   \put(155,0){$\iota_{n-1}$}  \put(135,55){$j_n$}
\put(135,10){\line(0,1){40}} \put(210,10){\line(1,0){75}} \put(280,0){$j_1$}
\put(235,10){\line(0,1){40}} \put(260,10){\line(0,1){40}}
\put(235,55){$j_3$}\put(260,55){$j_2$} \put(245,0){$\iota_2$}
\put(210,0){$\iota_3$}
\put(180,9){$\cdots$} \put(80,9){$\cdots$}
\end{picture}
\een
one reads off the following screening conditions
\bea
j_1+j_2-\iota_2&=&\rho_2-\sigma_2 t\nn
\iota_2+j_3-\iota_3&=&\rho_3-\sigma_3 t\nn
&\vdots&\nn
\iota_{n-1}+j_n-\iota_n&=&\rho_n-\sigma_n t\nn
&\vdots&\nn
\iota_{N-2}+j_{N-1}-j_N&=&\rho_{N-1}-\sigma_{N-1}t\nn
2j_i+1&=&r_i-s_it
\eea
with $\sigma_n,\rho_n$ non-negative integers, while the last line is the usual
parametrization of the weights. Following the procedure outlined in the
previous
section one may calculate the ghost field contribution to the correlator, while
the $\varphi$ part is a matter of standard computation. Let us summarize our
findings in a compact notation
\bea
M&=&\sum_{m=2}^{N-1}(\rho_m+\sigma_m)\nn
w_i&&i=1,..., M
\eea
$w_i$ collectively denote the positions of all screening charges. Furthermore
we introduce
\bea
k_i&=&\left\{\begin{array}{rl}
-1&i=1,...,\sum_m\rho_m\\
t&i=\sum_m\rho_m+1,...,M
\end{array}\right. \nn
B(w_i)&\equiv&\sum_{\ell=1}^{N-1}\frac{x_{\ell}/u_{\ell}}{w_i-z_\ell}
\label{B}
\eea
(here $x_1=0$).
Finally we may write down an integral representation of the $N$ point
conformal block
\ben
W_N=\oint\prod_{i=1}^M\frac{dw_i}{2\pi i}\oint\prod_{m=2}^{N-1}
  \frac{du_m}{2\pi i}W_N^{\beta\g}W_N^{\varphi}
\label{N}
\een
where
\bea
 W_N^{\beta\g}&=&\prod_{i=1}^M B(w_i)^{-k_i}
 \prod_{m=2}^{N-1}\Gamma(2j_m+1)u_m^{2j_m-1}e^{\frac{1}{u_m}}\nn
 W_N^{\varphi}&=&\prod_{m<n}(z_m-z_n)^{2j_mj_n/t}\prod_{i=1}^M\prod_{m=1}^{N-1}
(w_i-z_m)^{2k_ij_m/t}\prod_{i<j<M}(w_i-w_j)^{2k_ik_j/t}
\eea
This is the main result.

\section{The Knizhnik-Zamolodchikov equations}
\noindent
In ref. \cite{PRY1} several non-trivial consistency checks are
provided. Except for the case of the Knizhnik-Zamolodchikov (KZ) equations we
refer to this paper for a discussion of these, which include projective
invariance and a verification of the equivalence of the following two
correlators:
\bea
&&W^{(I)}_N(z_N=\infty,x_N=\infty,z_{N-1},x_{N-1},...,z_2,x_2,z_1=0,x_1=0) \nn
&=&\bra{j_N}{[}\phi_{j_{N-1}}(z_{N-1},x_{N-1}){]}^{j_N}_{\iota_{N-2}}...
{[}\phi_{j_2}(z_2,x_2){]}^{\iota_2}_{j_1}\ket{j_1}
\eea
and
\bea
&&W^{(II)}_N(z_N,x_N,z_{N-1},x_{N-1},...,z_2,x_2,z_1,x_1)\nn
&=&\bra{0}{[}\phi_{j_N}(z_N,x_N){]}^0_{j_N}
{[}\phi_{j_{N-1}}(z_{N-1},x_{N-1}){]}^{j_N}_{\iota_{N-2}}...\nn
&&...{[}\phi_{j_2}(z_2,x_2){]}^{\iota_2}_{j_1}
{[}\phi_{j_1}(z_1,x_1){]}^{j_1}_0 \ket{0}
\eea
The non-triviality of this check stems from the fact that the second
correlator involves more screening charges around the last field than the
first one.

In the formalism using the $x$ parameters, the KZ equations (which expresses
the decoupling of singular vectors) are the differential equations
\ben
\{t\pa_{z_{m_0}}+\sum_{m\neq m_0}
\frac{2D^a_{x_{m_0}}D^a_{x_m}}{z_m-z_{m_0}}\}W_N
=0
\label{KZ}
\een
where $z_{m_0}$ refers to the position of a selected field in the correlator.
What we want to check is that our final expression for the $N$ point function
(\ref{N}) fulfils the KZ equations {\it without} using any formal properties,
like associativity, of the algebra. This is then merely a consistency check of
our formalism. Let us define the "God given" function
\ben
G(w)=\frac{1}{w-z_{m_0}}\{D^+_{x_{m_0}}G^-(w)+2D^3_{x_{m_0}}G^3(w)+
D^-_{x_{m_0}}G^+(w)\}
\label{G}
\een
where the defining functions are (using $D_{B_i}\equiv\frac{\pa}{\pa B(w_i)}$)
\bea
G^+(w)&=&W_N^{\beta\g}W^\varphi_N\nn
G^3(w)&=&
   \left(B(w)\sum_{i=1}^M\frac{D_{B_i}}{w-w_i}+\sum_{i=1}^M\frac{k_i}{w-w_i}
 +\sum_{m=1}^{N-1}\frac{j_m}{w-z_m}\right)W_N^{\beta\g}W^\varphi_N\nn
G^-(w)&=&
 \left(-\sum_{i,j}B(w)\frac{D_{B_i}D_{B_j}}{(w-w_i)(w-w_j)}
+(t-2)\sum_i\frac{D_{B_i}}{(w-w_i)^2}\right.\nn
&-&\left.2\sum_{i,j}\frac{k_iD_{B_j}}{(w-w_i)(w-w_j)} -
2\sum_{m,j}\frac{j_mD_{B_j}}{(w-z_m)(w-w_j)}\right)W_N^{\beta\g}W^\varphi_N
\label{Gs}
\eea
(this is a slight abuse of notation since the right hand sides are supposed
to be integrated as in (\ref{N})).
On the left hand side one can show that $G(w)$ behaves like ${\cal O}(w^{-2})$,
which means that the total sum of pole residues must vanish. On the right hand
side one may carry out explicitly the computation of the pole residues, and
one then finds that the vanishing condition for the sum of these is exactly
the KZ equations. One might wonder why it is precisely this function
(\ref{G}) which apply for such an argumentation. The answer is quite simple,
because the expressions
\ben
G^a(w)=\br J^a(w){\cal O}\kt
\een
(where ${\cal O}$ is the collection of free field realizations of all the
chiral vertex operators and screening charges) appear in the usual proof of
the KZ equations, and it is indeed this way we have found (\ref{Gs}).

\section{Outlook}
\noindent
We believe that the techniques developed and results obtained in
the case of $SL(2)$
may be generalized to higher (super-)groups. Indeed for $SL(n)$
and other simple groups we have
managed to perform many of the generalizations, and in the case of $SL(3)$
we will hopefully soon be able to write down integral expressions for
at least three point functions  and determine the fusion rules \cite{PRY3}.
Then it should be possible also to generalize the discussion of Hamiltonian
reduction in our recent paper
\cite{PRY2} to $SL(3)$ and thereby shed some light on $W_3$
gravity. Generalizations to even higher (super-)groups might open up for
a treatment of more general non-critical string theories.

\vspace{4mm}\noindent
{\bf Acknowledgement}:
We are indebted to Anna Tollst\'en for discussions at the
early stages of this work.
Ming Yu would like to thank J.-B. Zuber and F.G. Malikov for private
communications. We are grateful to A.Ch. Ganchev and V.B. Petkova
for illuminating correspondence.
M.Y. also thanks the Danish Research Academy for financial support.

\end{document}